\documentclass[]{rsos}%%%%where rsos is the template name
\usepackage{algorithm, algpseudocode}

\begin{document}

%%%% Article title to be placed here
\title{Parameter estimation from aggregate observations: A Wasserstein distance based 
sequential Monte Carlo sampler}

\author{%%%% Author details
Chen Cheng$^{1}$, Linjie Wen$^{2}$ and Jinglai Li$^{3}$}

%%%%%%%%% Insert author address here
\address{$^{1}$School of Mathematical Sciences, \protect\\ Shanghai Jiao Tong University, Shanghai 200240, China\\
$^{2}$School of Earth and Space Sciences, \protect\\ Peking University, 5 Yiheyuan Rd, Beijing 100871, China\\
$^{3}$School of Mathematics, \protect\\ University of Birmingham, Edgbaston, Birmingham B15 2TT, UK}

%%%% Subject entries to be placed here %%%%
\subject{xxxxx, xxxxx, xxxx}

%%%% Keyword entries to be placed here %%%%

%%%% Insert corresponding author and its email address}
\corres{Jinglai Li\\
\email{j.li.10@bham.ac.uk}}

%%%% Abstract text to be placed here %%%%%%%%%%%%
\begin{abstract}
In this work we study systems consisting of a group of moving particles. 
In such systems, often some important parameters are unknown and have to be estimated from observed data. 
Such parameter estimation problems can often be solved via a Bayesian inference framework.
However in many practical problems, only data at the aggregate level is available
and as a result the likelihood function is not available, which poses challenge for Bayesian methods. In particular, we consider the situation where the distributions of the particles are observed. We propose a Wasserstein distance based sequential Monte Carlo sampler to solve the problem: the Wasserstein distance is used to measure the similarity between the observed
and the simulated particle distributions and the sequential Monte Carlo samplers is used 
to deal with the sequentially available observations. 
Two real-world examples are provided to demonstrate the performance of the proposed method.
\end{abstract}
%%%%%%%%%%%%%%%%%%%%%%%%%%%
\maketitle
%%%%%%%%%% Insert the texts which can accomdate on firstpage in the tag "fmtext" %%%%%

%$\begin{fmtext}
%\end{fmtext}
\section{Introduction}
In transportation science,
it is often of critical importance to study the collective behaviour of a 
group of moving objectives (referred to as particles hereafter).
A good example of such problems is the dynamics of a crowd of human pedestrians,
which has important applications in
urban planning and safety management~\cite{johansson2008crowd,Helbing2011}.  
Other examples of such systems include traffic flows~\cite{krauss1997metastable}, swarm robots~\cite{schranz2020swarm} and so on. 
Modeling these collective behaviors has 
attracted considerable attention in multiple disciplines,
and various models have been proposed in the past decades. 
Despite the modeling advances, some issues are yet to be solved. 
In particular,  certain key information in the system may be missing, 
and as a result some important  model parameters are not known to the practitioners.
Examples of such a situation may appear in pedestrian crowds~\cite{johansson2007specification}
and traffic flows \cite{hoogendoorn2010calibration,paz2015calibration}. 
In practice,  these model parameters are often estimated by fitting the real-world 
observation data into the mathematical models, often via a Bayesian framework. 

%%%%%%%%%%%%%%% End of first page %%%%%%%%%%%%%%%%%%%%%

The Bayesian parameter estimation is conceptually straightforward and has been used to estimate model
parameters in many related problems~\cite{gomes2019parameter,corbetta2015parameter,godel_bayesian_2022,makinoshima_crowd_2022}. 
However, it can be highly challenging to apply the Bayesian method to the microscopic models of such systems -- i.e., models that directly describe the dynamics of the individual particles in the system. 
A main difficulty is that, in order to conduct the standard Bayesian inference for a microscopic model, one needs to track each particle, 
which is extremely difficult or even impossible when the ensemble size is large. Instead, in reality, one often observes the aggregate data, 
that are collected at the ensemble level and do not characterise the state of each individual particle. As will be explained later, determining a suitable distribution for aggregate measures in this case can be challenging, potentially resulting in the unavailability of an analytically derived likelihood function. Additionally, the presence of unknown measurement noise further complicates the tractability of the likelihood. As a consequence, standard posterior computation methods such as Markov Chain Monte Carlo (MCMC) can not be used.
A large class of approximate Bayesian inference techniques have been developed to address such 
likelihood-free problems, such as the Approximate Bayesian Computation (ABC)~\cite{csillery2010approximate} and other methods, e.g., \cite{cranmer2020frontier}.
 These techniques have been used to conduct likelihood-free inference from aggregate observations. For instance, macroscopic observation data measuring the flow is used to calibrate a microscopic model for crowd dynamics using ABC methods~\cite{godel_bayesian_2022}.  

\begin{figure*}[!tb]
\centering
\includegraphics[width=\textwidth]{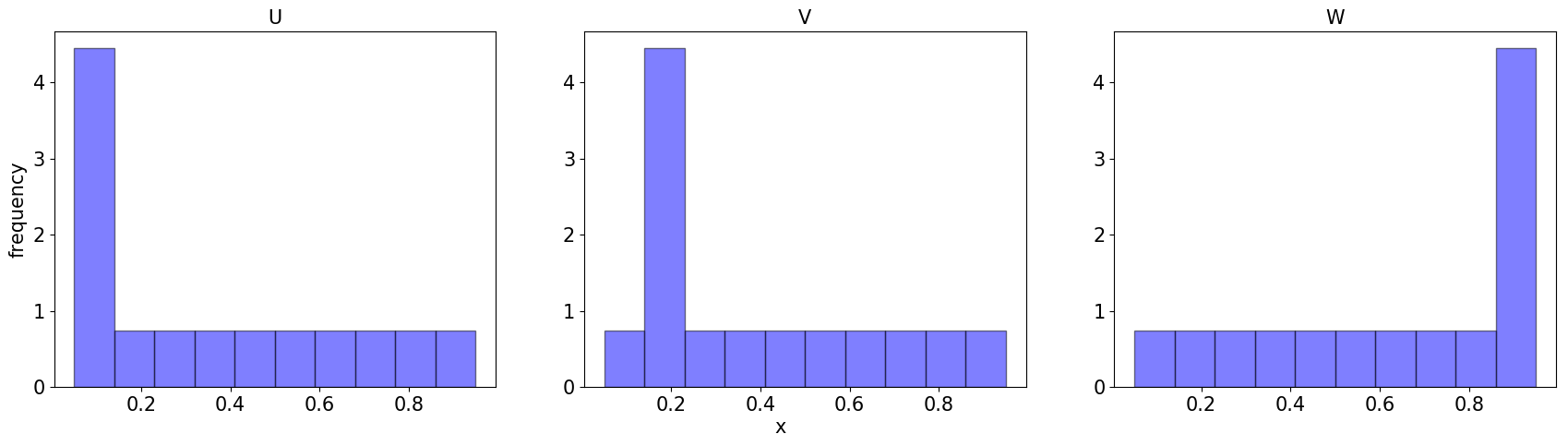}
\caption{An illustrative example of the TV distance and the Wasserstein distance where the histograms of three probability measures, $U$, $V$ and $W$, are shown. 
The TV distance between $U$ and $V$ and that between $U$ and $W$ are the same. The WD between $U$ and $V$ is 0.03, while the WD between $U$ and $W$ is 0.3. See the main text for more information.}
\label{fig.hist}
\end{figure*}

In the present work we consider a special case of aggregate observations -- namely it is the distribution of the particles that is observed. 
Our main purpose is to develop a generic method that can compute the posterior in such problems, by taking advantage of the rich literature of the likelihood-free inference. 
As in our problem of interest, usually the data are observed in a sequential manner, we choose to base our method on the sequential Monte Carlo sampler (SMCS)~\cite{del2006sequential}, an extension of the particle filter, for its ability to deal with sequentially available data. Some variants of SMCS can deal with likelihood-free inference~\cite{toni_approximate_2009} and these methods typically rely on the distance (e.g., the Euclidean distance) between the simulated data and the observed data. 
To this end a similar problem has been considered in the crowd dynamics context in \cite{makinoshima_crowd_2022}, where a crowd flow forecasting method based on particle filter is proposed to estimate both the crowd state and latent parameters from the aggregate density observation data.
In \cite{makinoshima_crowd_2022}, the Total-Variation (TV) distance is used to measure the similarity between the observed and the simulated distributions. 
However such distance measures may become ineffective when observation noise is present:
in Figure~\ref{fig.hist} we provide an example in which the TV distance between $U$ and $V$ and that between $U$ and $W$ are the same,
which makes it unsuited for measuring the distance between the observed and the simulated distributions. 
To address the issue, we propose to use the Wasserstein distance~\cite{rubner1998metric} as a distance measure between the simulated and the observed distributions. The Wasserstein distance~(WD) is a commonly used measure for similarity between two distributions, which loosely speaking, is the minimal cost for transforming one distribution into the other. In Figure~\ref{fig.hist}, the WD between $U$ and $V$ is computed as 0.03 while the WD between $U$ and $W$ is determined as 0.3. This discrepancy more reasonably characterizes the differences between these distributions and addresses the aforementioned issue with the TV distance. WD has been widely used in many applications in statistics and machine learning, such as~\cite{frogner2015learning,lee2018minimax},
as it has many desired properties. {Additionally, it has been shown to be effective in mitigating information loss caused by the utilization of summary statistics in basic ABC techniques~\cite{bernton_approximate_2019}.} In this work we develop a WD based SMCS method which can deal with the aforementioned problems where the observations are particle distributions (referred to as the aggregate data hereafter). 

The rest of the work is organised as follows. We first discuss the likelihood-free sequential inference problem in Section~\ref{sec:method},
and present the proposed Wasserstein-distance based SMCS in Section~\ref{sec:wd-smcs}.  Numerical examples are provided in Section~\ref{sec:examples} to demonstrate the performance of the proposed method. Finally Section~\ref{sec:conclusion} offers some conclusions.

\section{Likelihood-free Sequential Bayesian inference} \label{sec:method}

\subsection{Problem Setup}\label{sec:setup}
We start with a generic setup of the sequential Bayesian inference problem considered here.
{Suppose that we have measurements $y_1, ..., y_T$ (with $y_t \in \mathbb{R}^{n_y}$ for $t=1,...,T$) that are collected sequentially in time from a dynamical system model, 
and from the data we want to estimate the unknown parameter $\theta \in \mathbb{R}^{n_{\theta}}$}. 
In such problem, after collecting $t$ data points $y_1,...,y_t$, 
we can compute the posterior distribution of $\theta$ conditional on measurements $y_1,\,...,\,y_t$ in the following form (assuming all the measurements are independent conditional on $\theta$):
\begin{equation}\label{eq:posterior1}
\pi_t(\theta)=\pi\left(\theta | y_1, \ldots, y_t\right) \propto \pi_0(\theta) \prod_{i=1}^t \pi\left(y_i | \theta\right),
\end{equation}
where $\pi_0(\theta)$ is the prior distribution of $\theta$ that are usually specified by the users, 
or equivalently,
\begin{equation}\label{eq:posterior2}
\pi_t(\theta)\propto \pi_{t-1}(\theta)  \pi\left(y_t | \theta\right). 
\end{equation}
Our goal is to compute $\pi_T(\theta)$, i.e., the final posterior distribution after collecting all data points  $y_1, ..., y_T$. 

\subsection{Sequential Monte Carlo sampler}\label{sec:smcs}
Since the posterior distribution is typically not analytically available, 
a common practice is to characterise the distribution by drawing samples or realizations from it. In principle, the posterior $\pi_T(\theta)$ can be sampled with the standard  methods such as MCMC. 
However,  as is pointed out in \cite{fearnhead2013adaptive,zhou2016toward}, 
the posterior in Eq.~\eqref{eq:posterior1} poses challenges for usual MCMC methods especially 
when $T$ is large, as they need to query the likelihood function $\pi(\cdot \mid \theta)$ a large number of times.  
To this end, the SMCS method is particularly designed to exploit the sequential structure of such problems. 

SMCS is a sequential importance sampling method, 
and as such it generates weighted samples from the posterior distribution, where 
the samples represent possible values from the posterior, while the weights assign importance to each sample based on their contribution to capturing characteristics of the distribution.
We here describe SMCS in a recursive formulation. Given an arbitrary conditional distribution $L_{t-1}(\theta_{t-1}|\theta_t)$, one can construct the following joint distribution of $\theta_{t-1}$ and $\theta_t$:
\begin{equation}
p_t(\theta_{t-1},\theta_t)=\pi_{t}(\theta_t)L_{t-1}(\theta_{t-1}|\theta_t), 
\end{equation}
whose marginal distribution over $\theta_{t-1}$ is $\pi_t(\theta_t)$. Importance sampling~\cite{robert2013monte} is applied to draw samples for $p_t(\theta_{t-1},\theta_t)$, where the proposal distribution $q_t(\theta_{t-1},\theta_t)$ is constructed in the form of 
\begin{equation}
q_t(\theta_{t-1},\theta_t)=q_{t-1}(\theta_{t-1})K_t(\theta_t|\theta_{t-1}),
\end{equation}
from which we can directly draw samples $(\theta_{t-1}, \theta_t)$. Here $q_{t-1}(\theta_{t-1})$ is a marginal distribution and $K_t(\theta_t|\theta_{t-1})$ a conditional distribution.
The weights of samples are computed according to 
\begin{subequations}\label{eq:weight_all}
\begin{equation}
\begin{aligned}
\omega_t(\theta_{t-1:t}) &=\frac{p_t(\theta_{t-1},\theta_t)}{q_t(\theta_{t-1}, \theta_t)}=\frac{\pi_{t}(\theta_t)L_{t-1}(\theta_{t-1}|\theta_t)}{q_{t-1}(\theta_{t-1})K_t(\theta_t|\theta_{t-1})}\\
    &=\omega_{t-1}(\theta_{t-1})\alpha_t(\theta_{t-1},\theta_t),
\end{aligned}
\end{equation}
where 
\begin{equation}\label{eq:weight}
\omega_{t-1}(\theta_{t-1}) =\frac{\pi_{t-1}(\theta_{t-1})}{q_{t-1}(\theta_{t-1})}.
\end{equation}
is the weight at $t-1$ step, and 
\begin{equation}\label{eq:increment}
\alpha_t(\theta_{t-1},\theta_t)=\frac{\pi_{t}(\theta_t)L_{t-1}(\theta_{t-1}|\theta_t)}{\pi_{t-1}(\theta_{t-1})K_t(\theta_t|\theta_{t-1})},
\end{equation}
\end{subequations}
is the incremental weight at the $t$-th step. 
Now that we have joint samples $\{(\theta_{t-1}^j, \theta_t^j), \omega_t^j\}_{j=1}^N$ from the joint distribution $p_t(\theta_{t-1},\theta_t)$, it follows that samples $\{\theta_t^j, \omega_t^j\}_{j=1}^N$ correspond to the marginal distribution $\pi_t(\theta_t)$. The complete SMCS algorithm proceeds as the following: 
\begin{enumerate}
\item  let $t=0$, draw samples $\{\theta^j_{0}\}_{j=1}^N$ from $q_0(\cdot)$,
and compute $w^j_0=\pi_0(\theta^j_0)/q_0(\theta_0^j)$ for $j=1...N$;
\item let $t=t+1$; \label{st:t+1}
\item draw $\theta^j_{t}$ from $K_t(\cdot|\theta^j_{t-1})$ for each $j=1...N$;\label{st:sampling}
\item compute $w^{j}_{t}$ using Eq.~\eqref{eq:weight_all} for each $j=1...N$;
\item  go to step~\ref{st:t+1} if $t<T$.
\end{enumerate}
Finally the procedure returns a set of weighted samples $\{\theta_T^j, \omega_T^j\}_{j=1}^N$ from the posterior $\pi_T(\theta)$. Note that in SMCS algorithms, a resampling step is commonly used to alleviate the ``sample degeneracy'' issue~\cite{del2006sequential}. When the Effective Sample Size (ESS) falls below a specific threshold (usually less than half the total number of samples), resampling is performed on the proposed samples to mitigate this issue.

From the discussions above, we can see that in SMCS the forward kernel $K_{t}(\theta_t|\theta_{t-1})$ and the backward kernel $L_{t-1}(\theta_{t-1}|\theta_t)$ 
are critical for the performance of the method and therefore have to be chosen carefully. 
Here we adopt the MCMC moves developed in~\cite{del2006sequential}. The sampling step~\ref{st:sampling} in the SMCS algorithm with this forward kernel is constructed as follows. First we choose a proposal distribution $k(\theta_t|\theta_{t-1})$ and draw a sample $\theta^*_t$ from it with a sample $\theta_{t-1}^j$ from the previous iteration. 
Next the sample $\theta^*_t$ is accepted or rejected according to the following acceptance probability:
\begin{equation}\label{eq:acceptance}
a_t(\theta_t^* | \theta_{t-1}^j)=\min \left\{\frac{\pi_t(\theta_t^*)}{\pi_t(\theta_{t-1}^j)} \frac{k(\theta_{t-1}^j | \theta_t^*)}{k(\theta_t^* | \theta_{t-1}^j)}, 1\right\}.
\end{equation}
With this forward kernel, an approximate optimal backward kernel can be derived as:
\begin{equation}\label{eq:approx_opt}
L_{t-1}(\theta_{t-1}|\theta_t)=\frac{\pi_t(\theta_{t-1})K_t(\theta_t|\theta_{t-1})}{\pi_t(\theta_t)},
\end{equation}
where the detailed derivation can be found in \cite{del2006sequential} and is not repeated here.
With the choice of this approximate backward kernel, the incremental weight function $\alpha_t(\theta_{t-1},\theta_t)$ in Eq.~\eqref{eq:increment} reduces to:
\begin{equation}\label{eq:increment_new}
\alpha_t(\theta_{t-1},\theta_t)=\frac{\pi_t(\theta_{t-1})}{\pi_{t-1}(\theta_{t-1})},
\end{equation}
and when applied to the posterior distribution in Eq.~\eqref{eq:posterior2}, 
it becomes, 
\begin{equation}\label{eq:increment_final}
\alpha_t(\theta_{t-1},\theta_t)=\pi(y_t|\theta_{t-1}).
\end{equation}
Finally we provide some remarks on the SMCS method:
\begin{itemize}
    \item First we reinstate that SMCS is a special type of importance sampling (IS) approaches, as it draws samples from a proposal distribution instead of the actual posterior, while correcting for the bias by assigning proper weights to each samples. 
    Therefore, just like the standard IS, the method produces a set of sample-weight pairs which follow the posterior distribution. 
    We refer to \cite{robert2013monte} for more details of IS. 
    The main reason to use such a method is that it is challenging to draw samples directly from the posterior. SMCS is special as it constructs the joint target distribution as well as the joint proposal distribution in a special way -- using the forward and backward kernel functions. 
   \item The functions $K_t(\theta_t|\theta_{t-1})$ and $L_{t-1}(\theta_{t-1}|\theta_t)$ are both conditional distributions utilized within the SMCS method. They are not part of the inference problem and should not be regarded as likelihood functions. Instead, they can be understood as some auxiliary functions required by the SMCS method. More precisely they are used in constructing joint distributions along with a marginal distribution, and the terms ``forward'' and ``backward'' are associated with the dependence of their arguments.

\end{itemize}

\subsection{Likehood-free SMCS} \label{sec:abc}
It is easy to see that, to apply SMCS to the sequential inference problem described in Section~\ref{sec:setup}, one needs the knowledge of the likelihood function $\pi(\cdot|\theta)$.  
In many real world applications, including the problems of our interest in this paper, often the likelihood function is not explicitly available,
and instead there exists a simulation model that can generate synthetic observation data that distribute according to the likelihood function $\pi(\cdot|\theta)$,
given a parameter value $\theta$. 
In this case the approximate (or likelihood-free) inference methods can be used. 
While noting that many such approximate inference methods have been proposed, we focus on the one based on SMCS, which was modified from~\cite{calvet2011state}. 

Recall that, in the SMCS algorithm, the likelihood function is evaluated in two occasions:
the first is to calculate the acceptance probability in Eq.~\eqref{eq:acceptance}
and the second is to update the sample weights in Eq.~\eqref{eq:increment_final}.
The main idea of the approximate method is rather simple: 
one first generates a synthetic observation data $\hat{y}$ from the simulation model and approximates the value of the likelihood function
with a kernel function $H(\cdot,\cdot)$ that characterizes the similarity between $\hat{y}$ and the actual data $y$:
 \begin{equation}\label{eq.kernel}
\pi(y|\theta) \approx H(\hat{y}, y),
\end{equation}
where $\hat{y}$ is the synthetic observation data generated from $\pi(\hat{y}|\theta)$.
We reinstate here that the kernel function $H$ is chosen 
to characterise the similarity between $\hat{y}$ and $y$, 
and as such, the more similar $\hat{y}$ and $y$ are, the higher the likelihood function value is.

\section{Aggregate observations and the Wasserstein distance} \label{sec:wd-smcs}
The application problems of our interest are a special case of the sequential inference problems described in Section~\ref{sec:method}.
Specifically we consider the dynamics of an ensemble of $n$ particles (such as pedestrians or vehicles), which is described by 
a discrete-time dynamical system:

\begin{equation}
x_{t+1} = F_t(x_t;\theta),
\end{equation}
where $x_t=(x_{1t},...,x_{nt})$ with $x_{it}$ (for $i=1,...,n$) being the position of the $i$-th particle at time $t$, and $\theta$ is the unknown model parameter that we want to estimate. 
 
In the standard setup of such problems, one is able to observe the particle position $x_t$ 
at each time $t$, from which the parameter $\theta$ is inferred, and in this case,
the likelihood function is usually available.
However, in many real-world problems, especially when the number of particles is large,
it is not possible to  track and locate each particle,  
and instead it is often much easier to observe 
how the particles are distributed at a given time $t$, often referred to as \emph{aggregate observations}. 
Namely, at each $t=1,...,T$, we observe the distribution of $x_t$, denoted as $\rho_t(x)$,
and we aim to infer parameter $\theta$ from $\rho_t(x)$.
It should be clear that, since the relation between the parameter and the aggregate observations are complex, it is usually challenging to derive the likelihood function analytically. As such likelihood-free methods need to be used for such problems.
We adopt the likelihood-free SMCS method described in section~\ref{sec:abc} to solve the problem, and recall that,
in this framework, the key issue is to choose the kernel function $H(\cdot,\cdot)$ that measures the similarity between two data points that are essentially two distributions in the present problem. 
To this end, we choose to define the kernel function $H$ via the Wasserstein distance of two distributions. 
The Wasserstein distance, also known as the earth mover's distance (EMD), is a popular choice for measuring dissimilarity or distance between two distributions,  
and widely used in real-world applications.
Intuitively speaking the Wasserstein distance is calculated as the minimum ``cost'' of transforming one distribution into the other.
In what follows we discuss how to calculate the  Wasserstein distance between two discrete-valued distributions. Let 
$\rho_U=\{(u_1, p_1^u), ..., (u_m, p_m^u)\}, \rho_V=\{(v_1, p_1^v), ..., (v_n, p_n^v)\}$ be two discrete-valued distributions, where $p_i^u$ represents the density at the data point $u_i$. A function $c(u_i,v_j)$ is further defined as the cost for transporting a unit from $u_i\in U$ to $v_j \in V$, which is taken to be the Euclidean distance between $u_i$ and $v_j$ in this work.
To calculate the Wasserstein distance, we first need to compute  a ``transport map'' $\gamma(u_i, v_j)$ that minimize the overall cost
\begin{equation}
\sum_{i=1}^m \sum_{j=1}^n c(u_i, v_j) \gamma(u_i, v_j),
\end{equation}
subject to the following constraints:
\begin{equation}
\left\{\begin{array}{l}
\sum\limits_{j=1}^n \gamma(u_i, v_j)=p_i^u, \\
\sum\limits_{i=1}^m \gamma(u_i, v_j)=p_j^v, \\
\gamma \geq 0.
\end{array}\right.
\end{equation}
Once the optimal transport map $\gamma$ is found, the Wasserstein distance is defined as

\begin{equation}
D_W(\rho_U, \rho_V)=\frac{\sum_{i=1}^m \sum_{j=1}^n c(u_i, v_j)\gamma(u_i, v_j)}{\sum_{i=1}^m \sum_{j=1}^n \gamma(u_i, v_j)}.
\end{equation}
For more details of the Wasserstein distance we refer to~\cite{rubner1998metric} and 
the references therein. 
It follows that the kernel function can be defined as 
\begin{equation}\label{eq:emd}
H(\rho_U,\rho_V)= \frac{1}{\sqrt{2\pi}h} \exp{(-D_W(\rho_U,\rho_V)^2/2h^2)},
\end{equation}
where $h$ is the bandwidth parameter of the kernel. It is of essential importance 
to choose an appropriate value for $h$ as it directly affects the sensitivity of the kernel to the WD value. 
%Inadequate choices of $h$ can yield similar outputs for different WDs, making it challenging to distinguish samples based on their weights. 
The proper choices of $h$ is usually problem dependent, and more importantly, 
for a particular problem, as the WDs of samples can vary across iterations, a fixed value of $h$ may  not be suitable for all the iterations. As a result, we adopt an adaptive approach where $h$ is dynamically determined for each iteration. Namely we set $h$ to be the median WD among the samples. This adaptive strategy allows us to adjust the value of $h$ according to varying WDs encountered during the estimation process, ensuring an appropriate level of sensitivity for the kernel.
Finally we present the complete algorithm of the Wasserstein distance based SMCS in 
Alg.~\ref{alg:2}.

\begin{algorithm}[h]
\caption{Wasserstein distance based sequential Monte Carlo sampler (WD-SMCS)} 
\label{alg:2}
{\bf Input:} 
Number of samples $N$, Prior of parameters $\pi_0(\theta)$, observed distributions $y_{1:T}$, resampling threshold $\text{ESS}_{th}$
\smallskip
\begin{algorithmic}[1]
\State draw $N$ samples $\{\theta_0^j\}_{j=1}^N$ from prior $\pi_0(\theta)$ and initialize weights $\omega_0^j = \frac{1}{N}$;
\For{$t = 1, ..., T$}
\State draw candidate samples $\theta_t^{j*}\sim k(\cdot \mid \theta_{t-1}^j)$;
\State generate synthetic distributions $\hat{y}_{1:t}^j$ and $\hat{y}_{1:t}^{j*}$ with model parameter $\theta_{t-1}^j$ and $\theta_t^{j*}$;
\State compute pair-wise Wasserstein distance $D_W(\hat{y}_{1:t}^{j}, y_{1:t})$ and $D_W(\hat{y}_{1:t}^{j*}, y_{1:t})$;
\State compute approximated likelihood $\pi(y_{1:t} \mid \theta_t^{j*})$ and $\pi(y_{1:t} \mid \theta_{t-1}^j)$ via Eq.~\eqref{eq.kernel} and~\eqref{eq:emd};
\State compute acceptance $$a(\theta_t^{j*},\theta_{t-1}^j) = \min\{\frac{\pi_t(\theta_t^{j*})k(\theta_{t-1}^j \mid \theta_t^{j*})}{\pi_t(\theta_{t-1}^j)k(\theta_t^{j*} \mid \theta_{t-1}^j)}, 1\};$$
\State generate random $u\sim U(0,1)$, if $u<a(\theta_t^{j*},\theta_{t-1}^j)$, $\theta_t^j=\theta_t^{j*}$, else $\theta_t^j=\theta_{t-1}^{j}$;
\State update the weights with normalization $$\omega_t^j=\omega_{t-1}^j \pi(y_t \mid \theta_{t}^j);$$
\State calculate $\text{ESS}=1 / \sum_{j=1}^N (\omega_t^{j})^2$ and resample if $\text{ESS} < \text{ESS}_{th}$.
\EndFor
\end{algorithmic}
\smallskip
\end{algorithm}

\section{Numerical experiments}\label{sec:examples}
In this section, we consider two application problems, with which
we conduct numerical experiments to examine the performance of the proposed method. 
The first example is the classic social force model (SFM) for the crowd dynamics of pedestrians. 
The second is the intelligent driver model (IDM) for the collective behavior of traffic flow. 
We discuss the model details and the setup of the simulation scenarios in the following sections. 

\begin{figure}[!htb]
    \begin{minipage}[c]{.45\textwidth}
    \centering
    {\includegraphics[width=5.5cm, height=5cm]{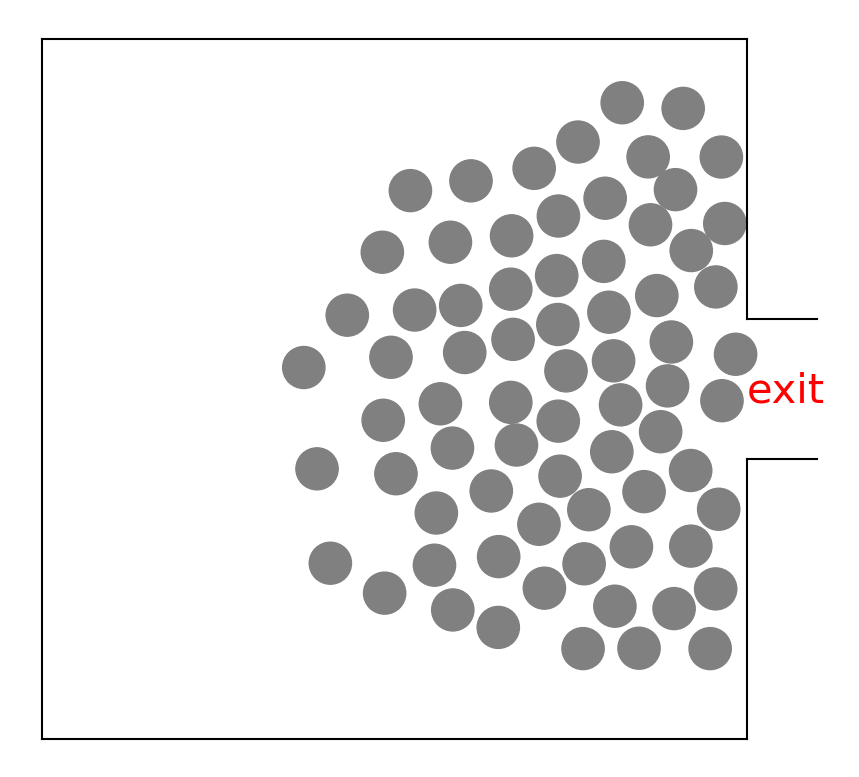}}
    \end{minipage}
    \begin{minipage}[c]{.45\textwidth}
    \centering
    {\includegraphics[width=7cm, height=2cm]{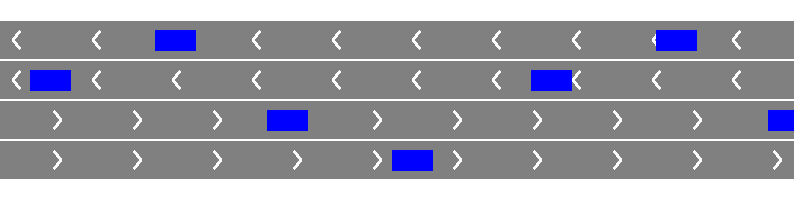}}
    \end{minipage}
\caption{Schematic plots of the simulation scenarios. (Left) Collective escape towards a single exit. Pedestrians are depicted by dots and walls (i.e., obstacles) are depicted by lines. All pedestrians are moving towards the center of the exit. (Right) Traffic flow on a four-lane highway. Vehicles are depicted by blue rectangles and white arrows on the roads indicate the driving direction of the vehicles.}
\label{fig:1}
\end{figure}

\subsection{The social force model}
Our first example is the dynamics of pedestrian crowds, and in particular 
we consider a typical scenario of collective escape towards a single exit (see Figure~\ref{fig:1}, Left). 
We adopt the social force model (SFM)~\cite{Helbing2000VICSEK} to describe the crowd behavior.
Simply speaking SFM assumes a mixture of physical and socio-psychological forces influencing the crowd behaviour 
by considering personal motivations and environmental
constraints. In this model, each pedestrian $i$ of mass $m_i$ and
velocity $\boldsymbol{v}_i$ tends to move by a
desired speed $v_i^p$ along a certain direction $\boldsymbol{e}_i^p$ during the acceleration
time $\tau_i$. The resulting personal desire force $\boldsymbol{F}_p$ is: 
\begin{eqnarray} \label{eq:1} 
\boldsymbol{F}_p = m_i\frac{v_i^p\boldsymbol{e}_i^p-\boldsymbol{v}_i}{\tau_i}. 
\end{eqnarray}
Furthermore, pedestrians psychologically tend to keep a social distance between each other 
and avoid hitting walls. This is modelled by introducing ``interaction force'' 
$\boldsymbol{f}_{ij}$ between pedestrians $i$ and $j$ and $\boldsymbol{f}_{iW}$ between 
pedestrian $i$ and the wall, respectively. The total interaction force is
\begin{eqnarray}\label{eq:2} 
\boldsymbol{F}_{int} = \sum\limits_{j(\neq i)}\boldsymbol{f}_{ij} + \sum\limits_{W}\boldsymbol{f}_{iW}. 
\end{eqnarray} 
Combining Eq.~\eqref{eq:1} and~\eqref{eq:2}, we obtain the acceleration equation
\begin{eqnarray} \label{eq:3} 
m_i\frac{\rm{d} \boldsymbol{v}_i}{\rm{d} t} = m_i\frac{v_i^p(t)\boldsymbol{e}_i^p(t) 
- \boldsymbol{v}_i(t)}{\tau_i} 
+ \sum\limits_{j(\neq i)}\boldsymbol{f}_{ij} 
+ \sum\limits_{W}\boldsymbol{f}_{iW}. 
\end{eqnarray} 
The position vector $\boldsymbol{r}_i(t)$ is updated by the velocity $\boldsymbol{v}_i(t)
  = \rm{d} \boldsymbol{r}_i/\rm{d} t$.

 The interaction force $\boldsymbol{f}_{ij}$ between pedestrian $i$ and $j$ is
 specified as follows.  With the distance $d_{ij} = ||\boldsymbol{r}_i - \boldsymbol{r}_j||$
 between the two pedestrians' centres of mass, the psychological tendency of
 pedestrian $i$ to stay away from pedestrian $j$ is described by a repulsive
 interaction force $A_i{\rm exp}[(r_{ij} - d_{ij})/B_i]\boldsymbol{n}_{ij}$, where
 $A_i$ and $B_i$ are constants, indicating the strength and the range of the
 interaction, and $\boldsymbol{n}_{ij} = (n_{ij}^1, n_{ij}^2) = (\boldsymbol{r}_i -
 \boldsymbol{r}_j)/d_{ij}$ is the normalised directional vector pointing from
 pedestrian $j$ to $i$. The pedestrians touch each other if their distance
 $d_{ij}$ is smaller than the sum $r_{ij} = r_i + r_j$ of their radius $r_i$ and
 $r_j$. In our model, we specify a uniform value for the size of each
 pedestrian (see Table~\ref{tab.1})~\cite{Helbing2000VICSEK}.  
 Inspired by granular interactions, two additional forces are included in the
 model, which are essential for understanding the particular effects in
 panicking crowds: a ``body force'' $k(r_{ij} - d_{ij})\boldsymbol{n}_{ij}$
 counteracting body compression and a ``sliding friction force'' $\kappa(r_{ij}
 - d_{ij})\Delta v_{ji}^t\boldsymbol{t}_{ij}$ impeding relative tangential motion, if
 pedestrians $i$ and $j$ are close enough. Here $\boldsymbol{t}_{ij} = (-n_{ij}^2, n_{ij}^1)$
 means the tangential direction and $\Delta v_{ji}^t = (\boldsymbol{v}_j -
 \boldsymbol{v}_i)\cdot\boldsymbol{t}_{ij}$ the tangential velocity difference, while $k$
 and $\kappa$ are large constants, representing the bump and the friction
 effect. In summary, the interaction force $\boldsymbol{f}_{ij}$ between pedestrians
 $i$ and $j$ is given by
 \begin{equation}\label{eq:4} 
 \begin{split}
\boldsymbol{f}_{ij} =& \{A_i{\rm exp}[(r_{ij} - d_{ij})/B_i] + k\mathbf{I}(r_{ij} - d_{ij})\}\boldsymbol{n}_{ij}\\ 
&+ \kappa \mathbf{I}(r_{ij} - d_{ij})\Delta v_{ji}^t\boldsymbol{t}_{ij},
\end{split}
\end{equation}
where the indicator function $\mathbf{I}(r_{ij} - d_{ij})$ is zero for
$r_{ij} - d_{ij} < 0$  and it is equal to $r_{ij} - d_{ij}$ otherwise.

The interaction with the walls is treated analogously. By denoting $d_{iW}$ as the distance 
to wall $W$, $\boldsymbol{n}_{iW}$ as the direction perpendicular to it, and $\boldsymbol{t}_{iW}$ as the 
direction tangential to it, we have
\begin{equation}\label{eq:5} 
\begin{split}
\boldsymbol{f}_{iW} =& \{A_i{\rm exp}[(r_i - d_{iW})/B_i] + kg(r_i - d_{iW})\}\boldsymbol{n}_{iW}\\ 
&- \kappa g(r_i - d_{iW})(\boldsymbol{v}_i\cdot\boldsymbol{t}_{iW})\boldsymbol{t}_{iW}.
\end{split}
\end{equation}

\begin{table}[h]
\small
\caption{List of parameter values in SFM and the three parameters to infer are in bold.}
\label{tab.1}
\begin{tabular}{lll}
 \hline
 Variable&Value&Description \\
 \hline
  $m$ &  $80\,\rm{kg}$  & mass of pedestrians \\
  $\mathbf{v^p}$ & $1.0\,\rm{m/s}$   & desired velocity \\
  $\tau$ & $0.5\,\rm{s}$  & acceleration time \\
  $r$ & $0.3\,\rm{m}$  & radius of pedestrians\\
  $\mathbf{A}$ & $2\times 10^3\,\rm{N}$  & interaction strength\\
  $\mathbf{B}$ & $0.08\,\rm{m}$ & interaction range\\
  $k$ & $1.2\times 10^5\,\rm{kg/s^2}$ & bump effect \\
  $\kappa$ & $2.4\times 10^5\,\rm{kg/(m\cdot s)}$ & friction effect \\
 \hline
\end{tabular}
\end{table}

With the SFM described above, we simulate the collective escape of pedestrians towards a single exit in a room. Initially, a total of 100 pedestrians are randomly distributed within a square room measuring 10 meters on each side. The room has an exit with a width of 2 meters. The parameter values used to simulate the data are given in Table~\ref{tab.1}, largely following~\cite{Helbing2000VICSEK}. 
We note here that in most of the real-world problems, the individuals are subject to different parameter values. To reduce the computational cost, we simplify the problem by assuming that the values for $m_i$, $v^p_i$, $\tau_i$, $r_i$, $A_i$ and $B_i$ are identical for all individuals, as are in~\cite{godel_bayesian_2022,makinoshima_crowd_2022}.
Among these parameters we assume that $v^p$, $A$ and $B$ are unknown and need to be inferred from the observation data. 
Namely, suppose we can observe the positions of pedestrians $x_t$ at different times $t^{obs} = t \times\Delta_t$ for $t = 1, \ldots, T$, and we aim to estimate the aforementioned parameters in the SFM with these sequential observations. The observation noise is assumed to be a zero-mean Guassian with standard deviation $\sigma$. At every observation time, noise is added to the actual positions of each particle and the underlying distribution of this noisy observation is estimated by its empirical densities.
We emphasize that we are able to observe the particle locations, but we are unable to track each particle. 
In this experiment, we take $\Delta_t = 0.1\rm{s}$ and $T = 30$ and the simulation time step in SFM is $dt = 0.001\rm{s}$. The prior distribution for $\theta = (A, B, v^p)^{\intercal}$ is taken to be uniform: $A\sim U[1200, 2200]$, $B\sim U[0.02, 0.2]$ and $v^p\sim U[0, 1.5]$. It is important to note that the SFM can encounter issues with unphysical parameter values that result in body compression and abrupt changes in velocities. Thus appropriate prior distributions of parameters are necessary to prevent such unrealistic simulations. Additionally, the simulation program includes an inspection procedure to identify and replace any unreasonable samples that may produce unrealistic signals during the simulation process. This approach enhances the stability of the simulation process and allows for more flexibility in setting parameter priors. We simulate the observation data with two different noise levels: $\sigma = 0.04$ and $\sigma = 0.1$, and use 500 samples in WD-SMCS to estimate parameters at each observation time. For each sample, the synthetic data $\hat{y}_t = \rho_t(x)$ is the approximated density of particle locations $x_t$ at observation time $t$. These particle locations are generated by the SFM simulation with the parameters assigned the values of the corresponding samples. Importantly, in the simulation step, no noise is added to the particle locations due to our assumption that we have no prior knowledge about the measurement noise.

\begin{figure*}[!htb]
\centering
\includegraphics[width=\textwidth]{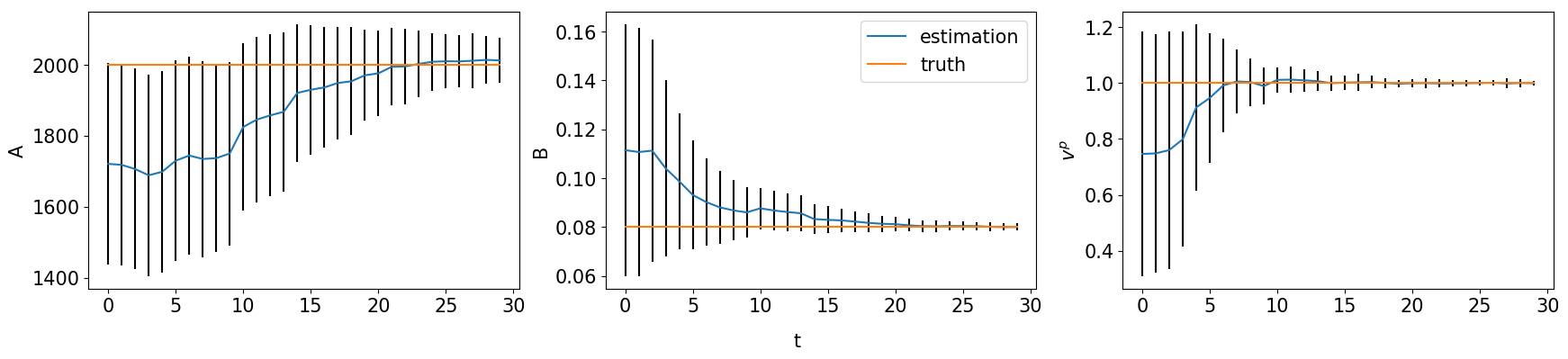}
\caption{($\sigma = 0.04$) Inference results of the SFM. Blue lines represent the average values of parameters estimated by 500 weighted samples and the standard deviations are shown as error bars. The horizontal lines represent the true parameter values.}
\label{fig.sfm_small}
\end{figure*}

\begin{figure*}[!htb]
\centering
\includegraphics[width=\textwidth]{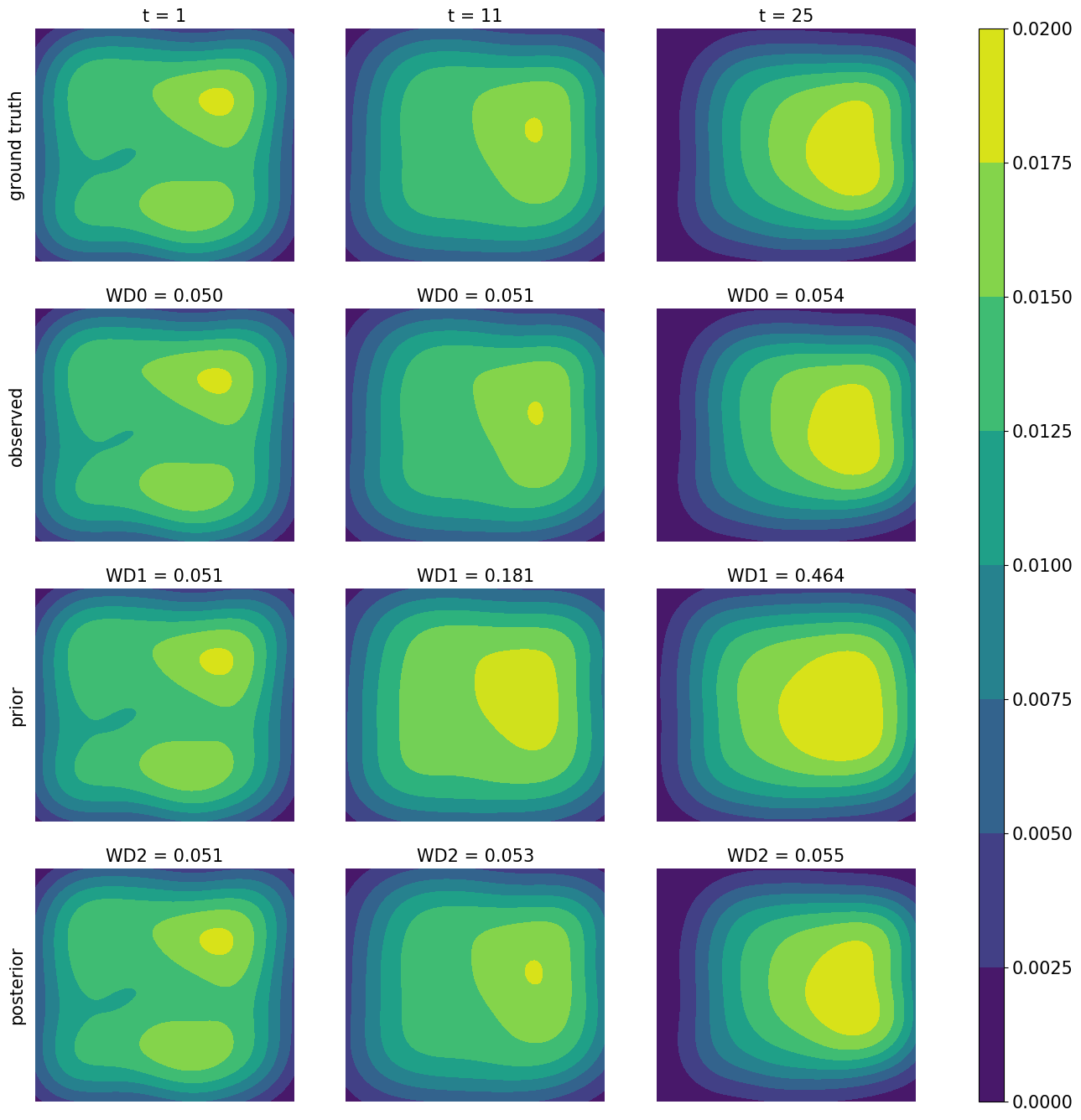}
\caption{($\sigma = 0.04$) Comparison among the ground truth (first row), the observation (second row), the prior prediction (third row) and the posterior prediction (last row) in terms of crowd density map at specific time points. The prior prediction is based on the mean values of parameter priors, while the posterior prediction is generated with the posterior means. WD0 is the WD between the ground truth and the observation. WD1 is the WD between the observed density and the prior-simulated one, and WD2 is that between the observed density and the posterior-generated one.}
\label{fig.density_small}
\end{figure*}

\begin{figure*}[!tb]
\centering
\includegraphics[width=\textwidth]{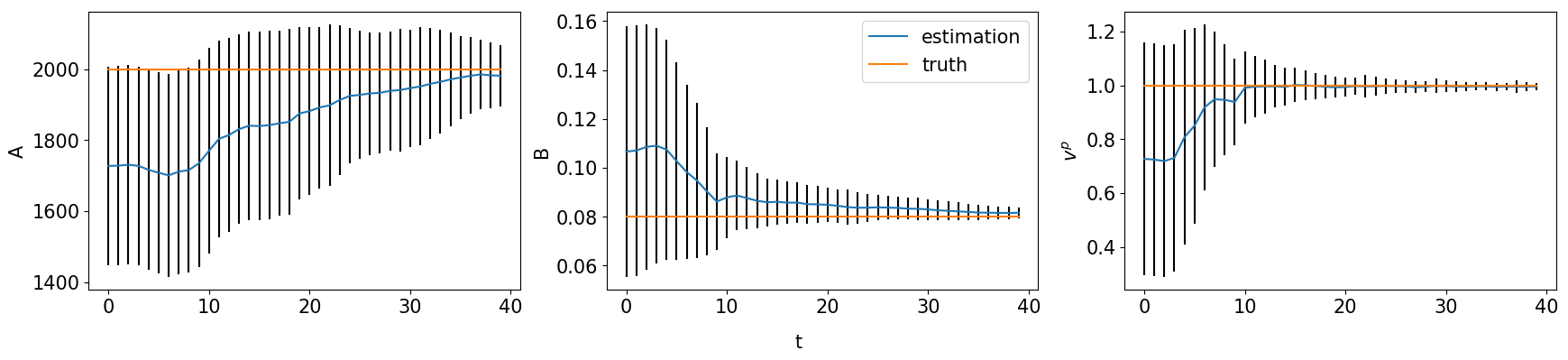}
\caption{($\sigma = 0.1$) Inference results of the SFM. Blue lines represent the average values of parameters estimated by 500 weighted samples and the standard deviations are shown as error bars. The horizontal lines represent the true parameter values.}
\label{fig.sfm_large}
\end{figure*}

\begin{figure*}[!tb]
\centering
\includegraphics[width=\textwidth]{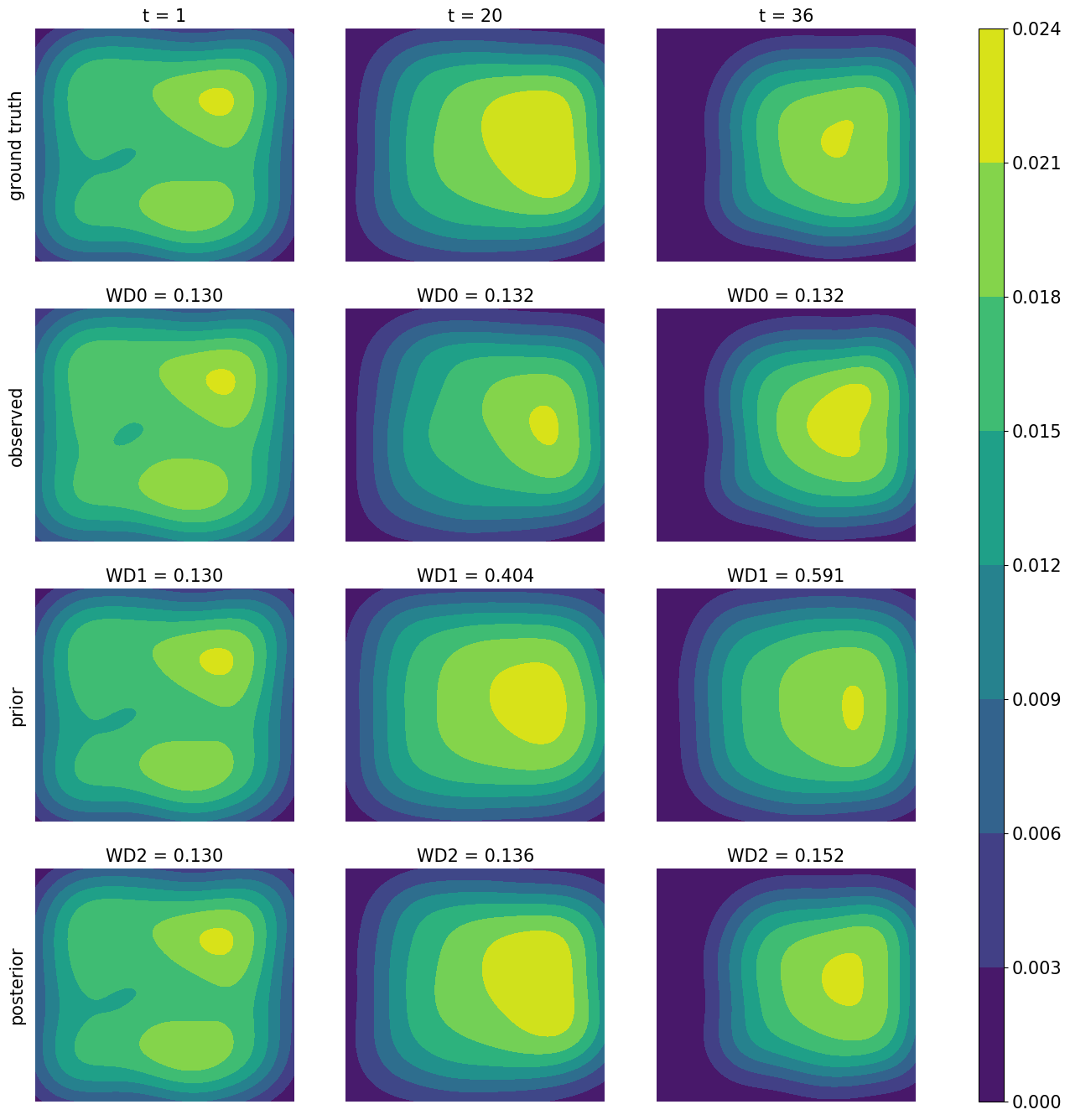}
\caption{($\sigma = 0.1$) Comparison among the ground truth (first row), the observation (second row), the prior prediction (third row) and the posterior prediction (last row) in terms of crowd density map at specific time points. The prior prediction is based on the mean values of parameter priors, while the posterior prediction is generated with the posterior means. WD0 is the WD between the ground truth and the observation. WD1 is the WD between the observed density and the prior-simulated one, and WD2 is that between the observed density and the posterior-generated one.}
\label{fig.density_large}
\end{figure*}

We first consider the small noise case where $\sigma=0.04$. 
In Figure~\ref{fig.sfm_small} we plot the posterior means and the standard deviations against the number of iterations for all the three parameters. 
As can be seen from the figure, the mean values of all three parameters converge within 25 sequential observations,
and the converged values are rather close to the ground truth, indicating good inference accuracy in the case when the observation noise is small. 
Recall that, a key element in the proposed method is that the simulated distribution should be close 
to the observed one. To facilitate the comparison of crowd distributions, we approximate the density map of particle locations using the kernel density estimation (KDE)~\cite{rosenblatt1956remarks} method.
In Figure~\ref{fig.density_small} we compare the observed particle density and that simulated with the posterior means (referred to
as the posterior-simulated density) at three time steps:
$t=1,\,11,\,25$;
 we also show the densities from the prior means (referred to as the prior-simulated density) and the ground truth as a reference. 
 For the two simulated densities we also calculate the WD between them to the observed one. It can be seen that as the iteration proceeds (and therefore more data is available) the density associated with the prior means deviates from the observation, the posterior prediction density is closer to the observation with a smaller WD, compared to the prior prediction density.
Next we consider the case of larger noise, i.e. $\sigma =0.1$.
As before we first plot the posterior means against the number of iterations in Figure~\ref{fig.sfm_large}.
One can see that, while the posterior means also approach to the true value as the iteration proceeds,
more  iterations (and hence more data) are required in order to accurately infer the parameters (especially $A$), due to the larger observation noise. 
We then show the same particle density plots as those in  Figure~\ref{fig.density_small}.
One can see here that Figure~\ref{fig.density_large} is qualitatively similar to Figure~\ref{fig.density_small}, which 
once again demonstrates that the proposed method can successfully obtain parameters values that can drive the particle density toward the observed one. 

\begin{figure*}[!tb]
\centering
\includegraphics[width=\textwidth]{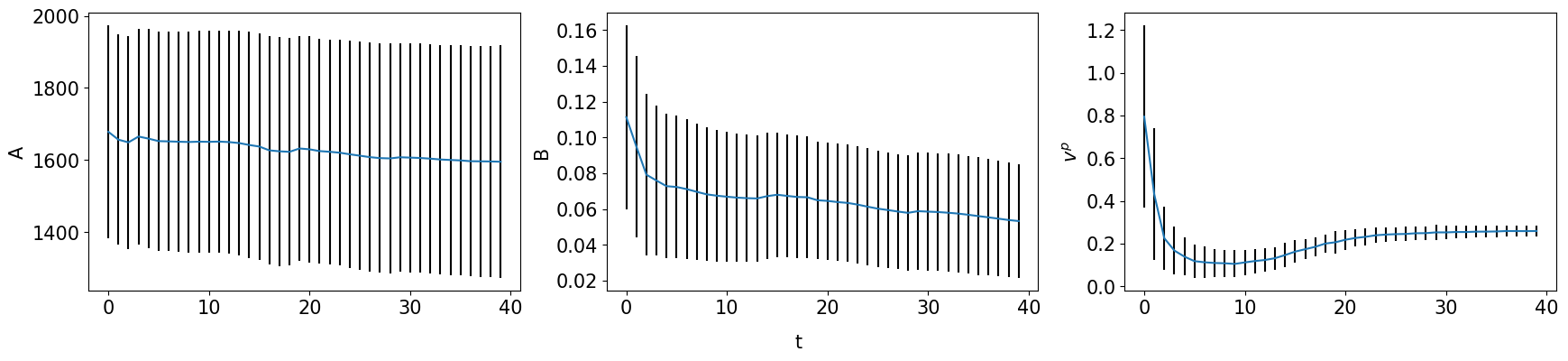}
\caption{Inference results of the real-world data. Blue lines represent the average values of parameters estimated by 500 weighted samples and the standard deviations are shown as error bars.}
\label{fig.sfm_real}
\end{figure*}

\begin{figure*}[!htb]
\centering
\includegraphics[width=\textwidth]{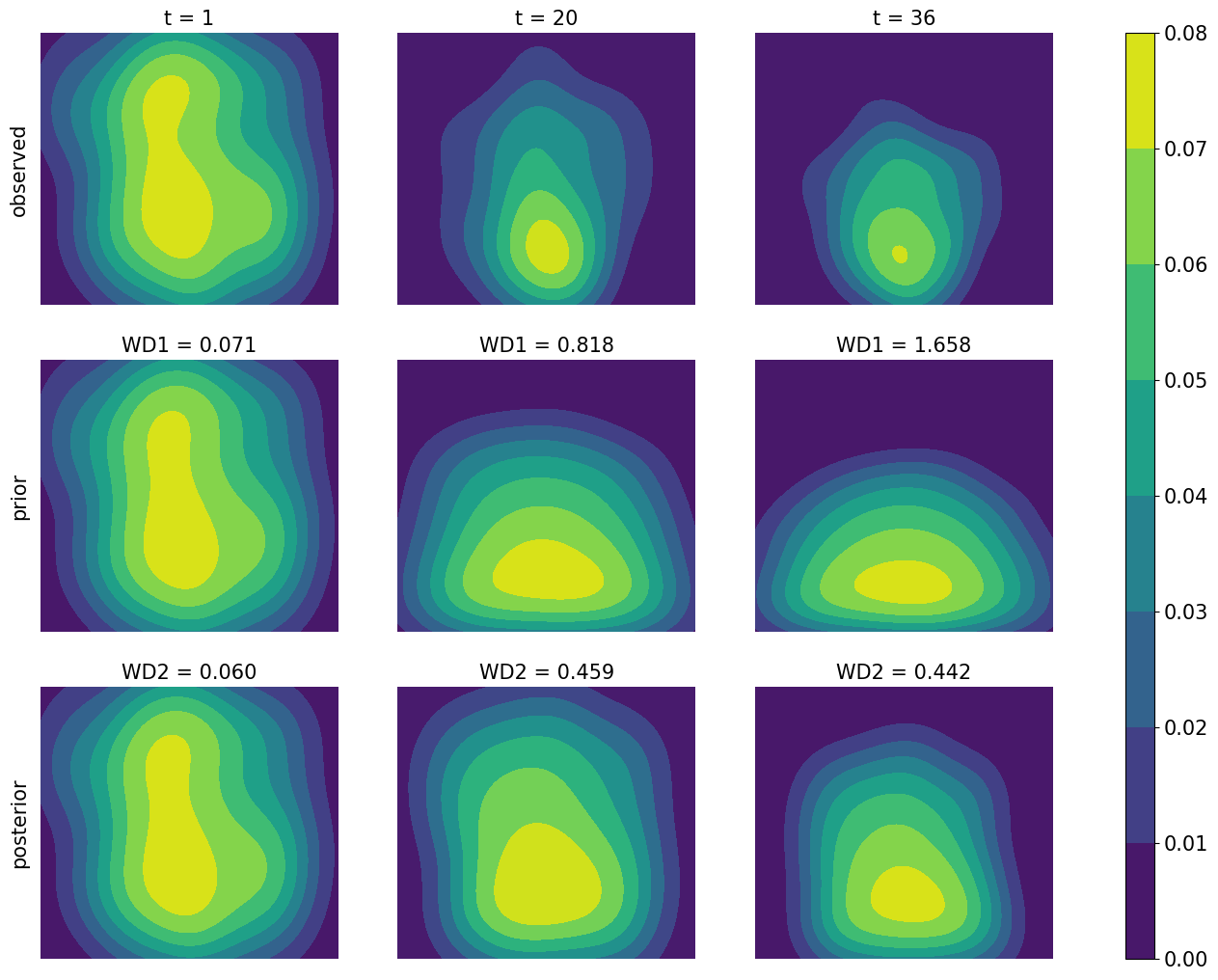}
\caption{Comparison among the observation (first row), the prior prediction (second row) and the posterior prediction (last row) in terms of crowd density map at specific time points. The prior prediction is based on the mean values of parameter priors, while the posterior prediction is generated with the posterior means. WD1 is the WD between the observed density and the prior-simulated one, and WD2 is that between the observed density and the posterior-generated one.}
\label{fig.density_real}
\end{figure*}

We conduct further testing of the proposed algorithm using real-world data. 
Specifically the experimental data~\cite{adrian2020crowds} is collected in a corridor scenario leading straight to an entrance gate, which is common for concerts or other events. Initially, there are 75 participants waiting in the corridor, which has a width of 5.6 meters. The exit is located along the negative y-axis and has a width of 0.5 meters. Once instructed, participants start moving towards the entrance gate, and their locations are recorded during the process. More detailed information about the experiment can be found in~\cite{adrian2020crowds}. We assume that the crowd dynamics in the corridor follow the SFM, and we use the observed distributions to infer the model's parameters. For this experiment, we take $\Delta_t = 0.2\rm{s}$ and $T = 40$. The parameters to be inferred and their priors are the same as before. In Figure~\ref{fig.sfm_real} we plot the posterior means and the standard deviations against the number of iterations for all the three parameters. From the figure, it can be observed that the mean value of the parameter $v^p$ converges within 30 sequential observations, while the mean values of the other two parameters, $A$ and $B$, remain highly variable. Given the unavailability of the actual model and parameter values, a reasonable practice is to assess whether the estimated model can capture the observed data patterns. Figure~\ref{fig.density_real} shows the comparison between densities. One can see that both the prior-generated and the posterior-generated densities present significant discrepancies compared to the observed density. However, the posterior density demonstrates a smaller WD to the observed density compared to the prior density. 
We note that the performance is ultimately limited by SFM, which may not accurately describe this particular scenario. Furthermore, the simplification of assuming identical parameters for each participant also affects the performance. Nevertheless, we can still extract some useful information from the data using the proposed method.

\subsection{The intelligent driver model}

Our second example is  the traffic on a four-lane highway (see Figure~\ref{fig:1}, Right), described by the intelligent driver model (IDM)~\cite{treiber2000congested}. 
IDM is a widely adopted car-following model in microscopic traffic simulation. In this type of models, as its name suggests, the $i$-th vehicle follows the $(i-1)$-th vehicle in front of it. The IDM assumes the acceleration of a follower $i$ is a continuous function of a set of inputs - the follower's current velocity $v_i$; the gap distance $s_i$ to the leader; and the relative velocity (approach rate) $\Delta v_i$ with respect to the preceding vehicle. The dynamics equation is defined as:
\begin{equation}
\dot{v}_i=a^{(i)}\left[1-\left(\frac{v_i}{v_0^{(i)}}\right)^\delta-\left(\frac{s^*\left(v_i, \Delta v_i\right)}{s_i}\right)^2\right] .
\end{equation}
in which
\begin{equation}
s_i=x_{i-1}-x_i-l^{(i)},
\end{equation}
\begin{equation}
\Delta v_i = v_i - v_{i-1},
\end{equation}
\begin{equation}
s^*(v, \Delta v)=s_0^{(i)}+T_s^{(i)} v+\frac{v \Delta v}{2 \sqrt{a^{(i)} b^{(i)}}}.
\end{equation}
This expression can be understood as an interpolation of a free road acceleration $a_f = a^{(i)}\left[1-(v_i/v_0^{(i)})^{\delta}\right]$ and an interaction deceleration $a_{int} = - a^{(i)}(s^*/s_i)^2$.
Other constant parameters include the maximum desired velocity $v_0^{(i)}$, jam distance $s_0^{(i)}$, maximum desired acceleration $a^{(i)}$ and deceleration $b^{(i)}$, safe time headway $T_s^{(i)}$ and configurable exponent parameter $\delta$, 
which controls the smoothness of the acceleration/deceleration. In this example, we study the case of identical vehicles whose model parameters $v_0^{(i)} = v_0$, $s_0^{(i)} = s_0$, $T_s^{(i)} = T_s$, $a^{(i)} = a$, $b^{(i)} = b$, $l^{(i)} = l$ are given in Table~\ref{tab.2}.

\begin{table}[h]
\caption{List of parameter values in IDM and the three parameters to infer are in bold.}
\label{tab.2}
\begin{tabular}{lll}
 \hline
 Variable&Value&Description \\
 \hline
  $\mathbf{v_0}$ &  $8.33\,\rm{m/s}$  & desired velocity \\
  $\mathbf{T_s}$ & $1.6\,\rm{s}$   & safe time headway \\
  $\mathbf{a}$ & $1.44\,\rm{m/s}$  & maximum acceleration \\
  $b$ & $4.61\,\rm{m/s}$  & maximum deceleration\\
  $\delta$ & $4$  & acceleration exponent\\
  $s_0$ & $2\,\rm{m}$ & jam distance\\
  $l$ & $5\,\rm{m}$ & vehicle length \\
 \hline
\end{tabular}
\end{table}

With the IDM described above, we simulate the traffic flow on a four-lane highway. Each lane spans 300 meters and starts empty. Vehicles arrive at a rate of three vehicles per second, which is intentionally chosen to ensure a sufficiently busy road, where interactions between vehicles occur frequently. Suppose that we are able to observe the vehicle locations $x_t$ at a sequence of discrete time points: $t^{obs} = t\times\Delta_t$ for $t = 1, \ldots, T$, and we want to estimate the parameters $\theta = (v_0, a, T_s)^{\intercal}$ in the IDM with these observed data with all other parameters known. The measurement noise is assumed to be a zero-mean Guassian with standard deviation $\sigma$. This noise is added to the actual positions of each vehicle at every observation time. In this experiment, we take $\Delta_t = 1.0\rm{s}$ and $T = 30$ and the simulation time step in IDM is $dt = 0.1\rm{s}$. The prior distribution  is taken to be uniform: $v_0\sim U[5.56, 22.22]$, $a\sim U[0.5, 5]$ and $T_s\sim U[0.5, 4]$. We simulate the observation data with two different noise levels: $\sigma = 0.1$ and $\sigma = 1.0$, and use 500 samples in WD-SMCS to estimate parameters at each observation time. For each sample, the synthetic data $\hat{y}_t = \rho_t(x)$ is the approximated density of vehicle locations $x_t$ at observation time $t$. These vehicle locations are generated by the IDM simulation with the parameters assigned the values of the corresponding samples.

\begin{figure*}[!htb]
\centering
\includegraphics[width=\textwidth]{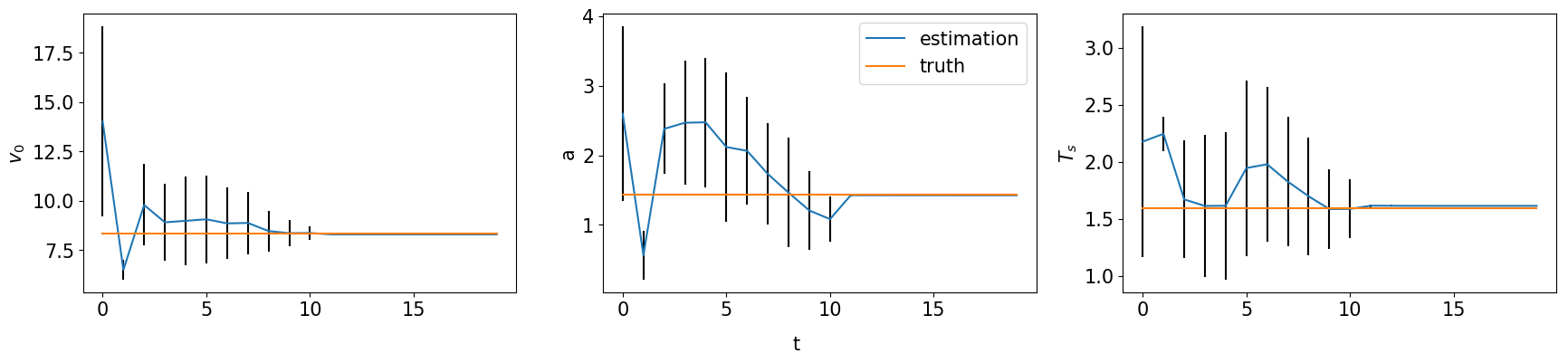}
\caption{( $\sigma=0.1$ ) Inference results of the IDM. Blue lines show the posterior means of the parameters and the posterior standard deviations are shown as error bars.
The horizontal lines represent the true parameter values.}
\label{fig.idm_small}
\end{figure*}

\begin{figure*}[!htb]
\centering
\includegraphics[width=\textwidth]{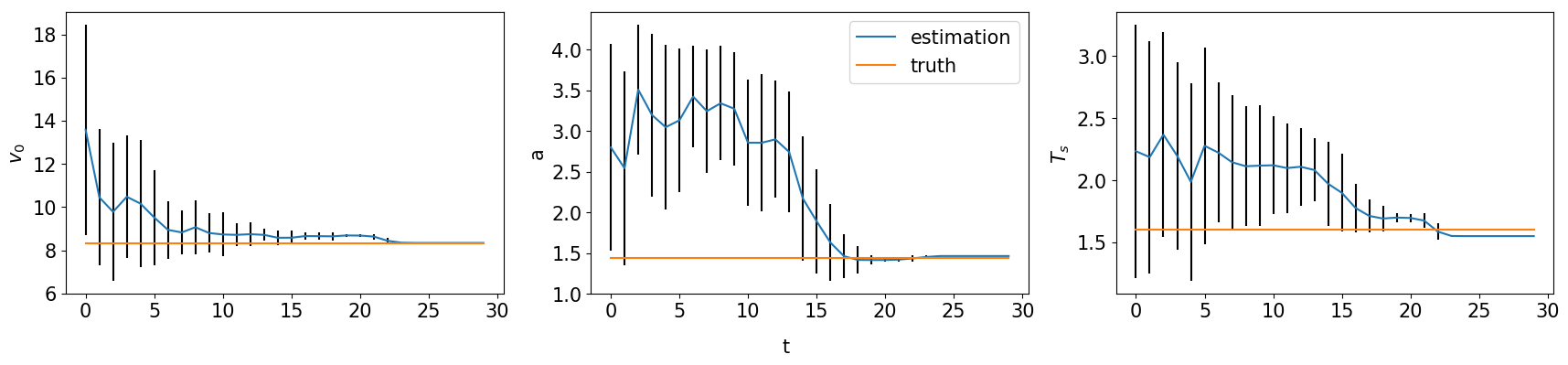}
\caption{( $\sigma=1.0$ ) Inference results of the IDM. Blue lines show the posterior means of the parameters and the posterior standard deviations are shown as error bars.
The horizontal lines represent the true parameter values.}
\label{fig.idm_large}
\end{figure*}

\begin{figure*}[!htb]
\centering
\includegraphics[width=\textwidth]{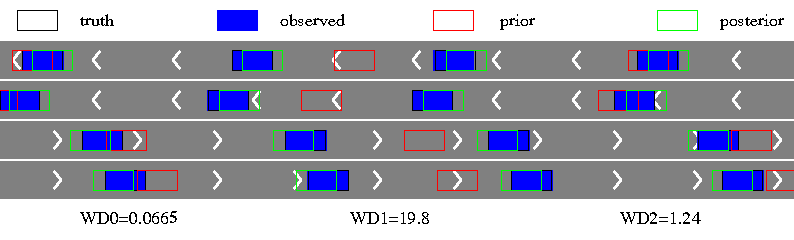}
\caption{( $\sigma=0.1$ ) The observed vehicle distribution (blue) compared against the ground truth (black), 
the posterior-simulated (green) and the prior-simulated (red) distributions over a specific interval on the highway. WD0 is the WD between the observed distribution and the ground truth, WD1 is the WD between the observed distribution and the prior-simulated one,
and WD2 is that between the observed distribution and the posterior-simulated one.}
\label{fig.highway_small}
\end{figure*}

\begin{figure*}[!htb]
\centering
\includegraphics[width=\textwidth]{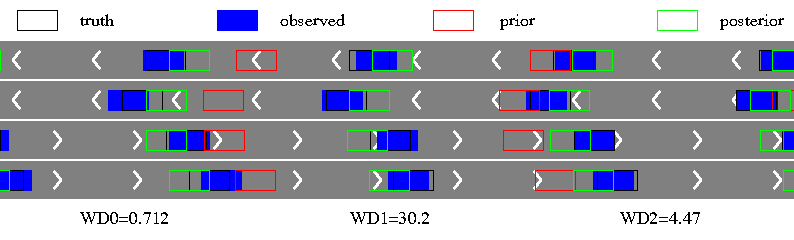}
\caption{( $\sigma=1.0$ ) The observed vehicle distribution (blue) compared against the ground truth (black), 
the posterior-simulated (green) and the prior-simulated (red) distributions over a specific interval on the highway. WD0 is the WD between the observed distribution and the ground truth, WD1 is the WD between the observed distribution and the prior-simulated one,
and WD2 is that between the observed distribution and the posterior-simulated one.}
\label{fig.highway_large}
\end{figure*}

We first consider the small noise case with $\sigma=0.1$ and plot the posterior means and standard deviations of the parameters against the number of iterations in Figure~\ref{fig.idm_small}. 
As can be seen from  the figure, the mean estimated values of all three parameters converge to almost the ground truth within 15 sequential observations, which indicates the accurate inference in the case when the observation noise is small.  The plots for the large noise case (i.e. $\sigma = 1.0$) are shown in Figure~\ref{fig.idm_large},
where we can see that it requires more  data points to infer the parameters (especially $a$ and $T_s$), but rather accurate results can still be obtained within 25 sequential observations. We also provide comparison of the vehicle distributions in Figure~\ref{fig.highway_small} and Figure~\ref{fig.highway_large} for the small and the large noise respectively. In this example, comparing the distribution of vehicles based on their locations is more straightforward and visually interpretable due to the clear separation by lanes. Additionally, to accommodate the space limitations in the paper, we plot a subset of vehicle locations over a specific interval on the highway, covering approximately 100 meters.
Specifically the figures show the observed vehicle distributions (blue), the posterior-simulated (green) and the prior-simulated (red) vehicle distributions. Visually in both figures the posterior-simulated distribution is rather close to the observation,  while the prior-simulated one clearly deviates. Quantitatively the WD between the posterior-simulated distribution and the observed one is much smaller than that between the prior-simulated and the observed ones: they are 19.8 versus 1.24 in the small noise case and 30.2 versus 4.47 in the large noise case.

\section{Conclusion}\label{sec:conclusion}
In summary, we consider in this work the problem of estimating parameters in 
a many-particle systems. In particular we aim to address the issue that in such problems 
it is often possible to observe the distributions of the particles rather than the trace of each individual particle. In this case the likelihood function is not available and we adopt a likelihood-free SMCS method where the similarity (distance) between the observed data and 
the simulated data need to be characterised. 
Since the observed data is actually a distribution, we propose to use the Wasserstein distance to measure the similarity between the observed and the simulated data. 
Numerical experiments are also provided to demonstrate the performance of the proposed method. 
Our belief is that the proposed method has potential applications in a variety of real-world problems that involve the dynamics of multiple particles, particularly in the field of transport science. We intend to explore these potential applications in future research. 

\vskip.5pc

\enlargethispage{10pt}

% \ack{Insert acknowledgment text here.}

%%%%%%%%%% Insert bibliography here %%%%%%%%%%%%%%

% \bibliographystyle{RS}
% \bibliography{refs}

\end{document}